\documentclass[10pt,journal,twocolumn,twoside]{IEEEtran} 
\usepackage[bottom=0.950in,top=0.600in,left=0.610in,right=0.610in]{geometry}

\usepackage{xpatch}
\usepackage{graphicx}
\usepackage{subfigure}
\usepackage{epstopdf}
\usepackage{float}
\usepackage{algorithmic}
\usepackage{array}
\usepackage{amsmath}
\usepackage{amssymb}
\usepackage{bm}
\usepackage{mdwmath}
\usepackage[table]{xcolor}
\usepackage{makecell}
\usepackage{eqparbox}
\usepackage{stfloats}
\usepackage{fixltx2e}
\usepackage{tabularx}
\usepackage{cases} 
\usepackage{booktabs}
\usepackage{threeparttable}
\usepackage{xcolor}
\usepackage{makecell}
\usepackage{multirow}

\usepackage[boxed,ruled,commentsnumbered]{algorithm2e}
\usepackage{url}
\usepackage{cite}
\allowdisplaybreaks[4]

\makeatletter

\newcommand{\Rmnum}[1]{\expandafter\@slowromancap\romannumeral #1@}
\makeatother

\makeatletter
\renewcommand*{\@opargbegintheorem}[3]{\trivlist
      \item[\hskip \labelsep{\bfseries #1\ #2}] \textbf{(#3):}\ }
\makeatother     
\floatname{algorithm}{Procedure}

\begin{document}

\makeatletter
\def\changeBibColor#1{%
  \in@{#1}{}
  \ifin@\color{red}\else\normalcolor\fi
}
 
\xpatchcmd\@bibitem
  {\item}
  {\changeBibColor{#1}\item}
  {}{\fail}
 
\xpatchcmd\@lbibitem
  {\item}
  {\changeBibColor{#2}\item}
  {}{\fail}
\makeatother

\title{Benchmarking Semantic Communications for Image Transmission Over MIMO Interference Channels}

\author{Yanhu Wang, Shuaishuai Guo, \emph{Senior Member}, \emph{IEEE},  Anming Dong, and Hui Zhao
\thanks{

Yanhu Wang, and Shuaishuai Guo are with School of Control Science and Engineering, and also with Shandong Key Laboratory of Wireless Communication Technologies, Shandong University, Jinan, 250061, China (e-mail: yh-wang@mail.sdu.edu.cn, shuaishuai\textunderscore guo@sdu.edu.cn).

Anming Dong is with the School of Computer Science and Technology, Qilu University of Technology,
Jinan, 250353, China (e-mail: anmingdong@qlu.edu.cn).

Hui Zhao is with the Communication Systems Department, EURECOM,
06410 Sophia Antipolis, France (e-mail: hui.zhao@eurecom.fr).
}
}

\maketitle

\begin{abstract}
Semantic communications offer promising prospects for enhancing data transmission efficiency. 
However, existing schemes have predominantly concentrated on point-to-point transmissions. In this paper, we aim to investigate the validity of this claim in interference scenarios compared to baseline approaches. Specifically, our focus is on general multiple-input multiple-output (MIMO) interference channels, where we propose an interference-robust semantic communication (IRSC) scheme.
This scheme involves the development of transceivers based on neural networks (NNs), which integrate channel state information (CSI) either solely at the receiver or at both transmitter and receiver ends.
Moreover, we establish a composite loss function for training IRSC transceivers, along with a dynamic mechanism for updating the weights of various components in the loss function to enhance system fairness among users.  Experimental results demonstrate that the proposed IRSC scheme effectively learns to mitigate interference and outperforms baseline approaches, particularly in low signal-to-noise (SNR) regimes.

\end{abstract}


\IEEEpeerreviewmaketitle

\section{Introduction}
\IEEEPARstart{R}{ecent} times have witnessed a resurgence of interest in semantic communications\cite{10012981}, propelled by the expanding realm of intelligent applications such as augmented reality/virtual reality (AR/VR)\cite{10293197}.
These advanced applications impose stringent requirements on communication services, prompting the exploration of novel theories and technologies in wireless networks.
Semantic communication, as a new paradigm, aims to convey the meaning of data rather than focusing solely on the precise transmission of individual symbols\cite{10177738}.
This approach holds significant potential to reduce data traffic, thereby greatly improving communication efficiency and effectively supporting various intelligent applications.

With advancements in deep learning (DL), particularly in natural language processing (NLP) and computer vision (CV), significant research efforts have been dedicated to exploring the domain of semantic communications\cite{bourtsoulatze2019deep,10431795,10458014}.
Neural networks (NNs) are employed in \cite{bourtsoulatze2019deep,10431795,10458014} to construct semantic encoder/decoder models, which are subsequently trained on large-scale datasets to acquire the ability to extract and interpret semantic information.
These NNs-based semantic communications can handle various types of data, including images\cite{10483054}, speech\cite{9953316} and text\cite{Xie2021}.
When compared to conventional communication systems based on source-channel separation, they demonstrate superior performance at the same transmission rate.
However, most existing NNs-based works have focused on simplistic single-user channel models, such as additive white Gaussian noise (AWGN) channels or fading channels, which do not adequately address interference issues.
Interference is a common challenge in wireless communications, particularly for users located at the edge of cell networks.
Thus, the question of how to effectively integrate semantic communication in environments with interference remains unanswered.

Introducing unexpected interference to semantic communications may lead to erroneous data recovery.
Real-time monitoring of the communication environment and dynamically adjusting transmission power can reduce interference.
However, such adjustments may compromise communication efficiency and reliability.
To tackle these challenges, we propose a NNs-based multiple-antenna transceiver design to handle interference.
Specifically, we introduce an interference-robust semantic communication (IRSC) design over multiple-input multiple-output (MIMO) interference channels.
The IRSC scheme utilizes NNs for transceiver design and integrates channel state information (CSI).
We explore two variations in our design: one with CSI provided only at the receiver end and another with CSI available at both transmitter and receiver ends.
The transceivers are trained using a composite loss function. Additionally, we develop a dynamic mechanism to adjust the weights of various components in the loss function, thereby enhancing system fairness among users. The results demonstrate the effectiveness of the proposed IRSC scheme in mitigating interference. Compared to benchmark schemes, our approach performs well in low SNR regimes.



\section{System Model}

\begin{figure}
       \centering   
       \includegraphics[width=1\linewidth]{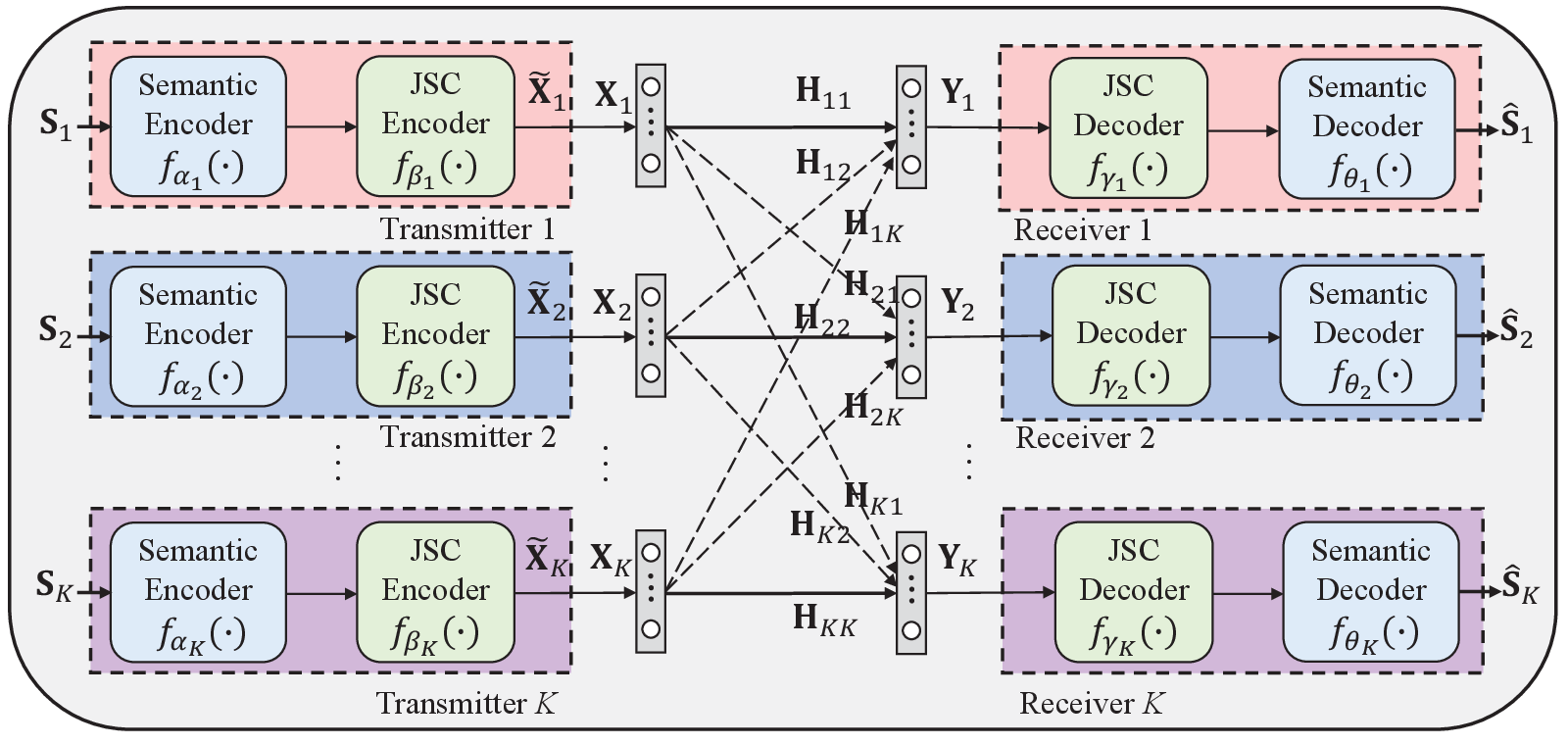}      
       \caption{Semantic communication system model over MIMO interference channels.
        The direct link is depicted by a solid line, and the interfering link is depicted by a dashed line.}
    \label{fig1}
\end{figure}

As depicted in Fig. \ref{fig1}, we consider a system comprising $K$ transmitter-receiver pairs, all of which operate over a shared physical channel. Without loss of generality, we presume that each transmitter is equipped with $N_t$ antennas, and each receiver with $N_r$ antennas.
In this setup, transmitters send semantic information to their corresponding receivers. However, each receiver can also accidentally receive semantic information from other transmitters, causing interference between users, as illustrated by the dashed lines in Fig. \ref{fig1}.

\subsection{Transmitter}
Let $\mathbf{S}_k \in \mathbb{R}^n$ represent the image input of user $k$, where $n$ denotes the dimension of the input images and $k=1,\cdots,K$.
As shown in Fig. \ref{fig1}, the transmitter consists
of semantic encoder and joint source-channel (JSC) encoder.
The semantic encoder extracts semantic information from $\mathbf{S}_k$ and then the JSC encoder maps them to symbols $\widetilde{\mathbf{X}}_k \in \mathbb{C}^{N_tN_B}$ to ensure reliable transmission over the channel.
Here, $N_tN_B$ denotes the number of complex-valued symbols.
This encoding process can be represented as follows:
\begin{equation}
\label{enc1}
\widetilde{\mathbf{X}}_k = f_{\boldsymbol{\mathit{\beta}}_k}(f_{\boldsymbol{\mathit{\alpha}}_k}(\mathbf{S}_k)),
\end{equation}
where $f_{\boldsymbol{\mathit{\alpha}}_k}(\cdot)$ and $f_{\boldsymbol{\mathit{\beta}}_k}(\cdot)$ represent the semantic encoder function and JSC encoder function for the transmitter $k$, parameterized by $\boldsymbol{\mathit{\alpha}}_k$ and $\boldsymbol{\mathit{\beta}}_k$ respectively.
The symbols $\widetilde{\mathbf{X}}_k$ are normalized and reshaped before transmission according to
\begin{equation}
\label{ap}
\mathbf{X}_k = {\rm Reshape} \left(\sqrt{PN_tN_B}\frac{\widetilde{\mathbf{X}}_k}{\left \|\widetilde{\mathbf{X}}_k \right\|_2} \right),
\end{equation}
where $P$ is the average transmit power constraint;
$\rm{Reshape}(\cdot)$ is a function that changes the shape of input; and
$\mathbf{X}_k \in \mathbb{C}^{N_t\times N_B}$ is the signal matrix transmitted by transmitter $k$.

\subsection{Communication Model}
The received symbol matrix at the downlink user $k$ can be given as
\begin{equation}
\label{chn}
\mathbf{Y}_k = \mathbf{H}_{kk} \mathbf{X}_k
+ \sum_{j=1, j\neq k}^K \mathbf{H}_{kj} \mathbf{X}_j + \mathbf{N}_k,
\end{equation}
where $\mathbf{Y}_k \in \mathbb{C}^{N_r\times N_B}$.
The downlink channel $\mathbf{H}$ is flat Rayleigh fading,
and $\mathbf{H}_{kj} \in \mathbb{C}^{N_r\times N_t} $ stands for the channel matrix from the $j$-th transmitter to the $k$-th receiver.
$\mathbf{N}_k\in\mathbb{C}^{N_r\times N_B}$ is the noise matrices with each entry obeying an independent identical zero-mean  Gaussian distribution with variance $\sigma^2$.
In \eqref{chn}, the first term on the right side of the equation is the expected signal, the second term is the interference between users, and the last term is the additive Gaussian noise.

\subsection{Receiver}
Similar to the transmitter, the receiver also comprises two parts, namely a JSC decoder and a semantic decoder.
JSC decoder is used to recover the transmitted symbols, and the semantic decoder is used to recover the transmitted images.
The decoding process at the receiver $k$ can be written as
\begin{equation}
\label{dec1}
\hat{\mathbf{S}}_k = f_{\bm{\theta}_k}(f_{\bm{\gamma}_k}(\mathbf{Y}_k)),
\end{equation}
where $\hat{\mathbf{S}}_k \in \mathbb{R}^n$ is the target image of user $k$;
$f_{\bm{\gamma}_k}(\cdot)$ and $f_{\bm{\theta}_k}(\cdot)$ are the JSC decoder function and semantic decoder function for the receiver $k$, parameterized by $\bm{\gamma}_k$ and $\bm{\theta}_k$ respectively.

\subsection{Problem Description}
The interference channels considered in this work may disrupt the transmission of semantic information and distort the intended meaning of data, thus affecting the overall efficiency and reliability of the semantic communication system.
We aim to maximize resistance to interference channels for accurate image transmission.
This problem can be formulated as
\begin{equation}
\begin{split}
\label{P1}
(\bm{\alpha}_k^*, \bm{\beta}_k^*, \bm{\gamma}_k^*, \bm{\theta}_k^*) &= \underset{\bm{\alpha}_k, \bm{\beta}_k, \bm{\gamma}_k, \bm{\theta}_k} {\rm argmin} \mathcal{L}_{MSE}(\bm{\alpha}_k, \bm{\beta}_k, \bm{\gamma}_k, \bm{\theta}_k) 
\\&=\underset{\bm{\alpha}_k, \bm{\beta}_k, \bm{\gamma}_k, \bm{\theta}_k} {\rm argmin}  \frac{1}{M} \sum^M_i (\mathbf{S}_{k,i}-\hat{\mathbf{S}}_{k,i})^2, 
\end{split}
\end{equation}
where the function $\mathcal{L}_{MSE}(\cdot)$ represents MSE between  $\mathbf{S}_k$ and $\hat{\mathbf{S}}_k$.
$M$ denotes the number of samples.

\begin{figure*}
       \centering   
       \includegraphics[width=0.84\linewidth]{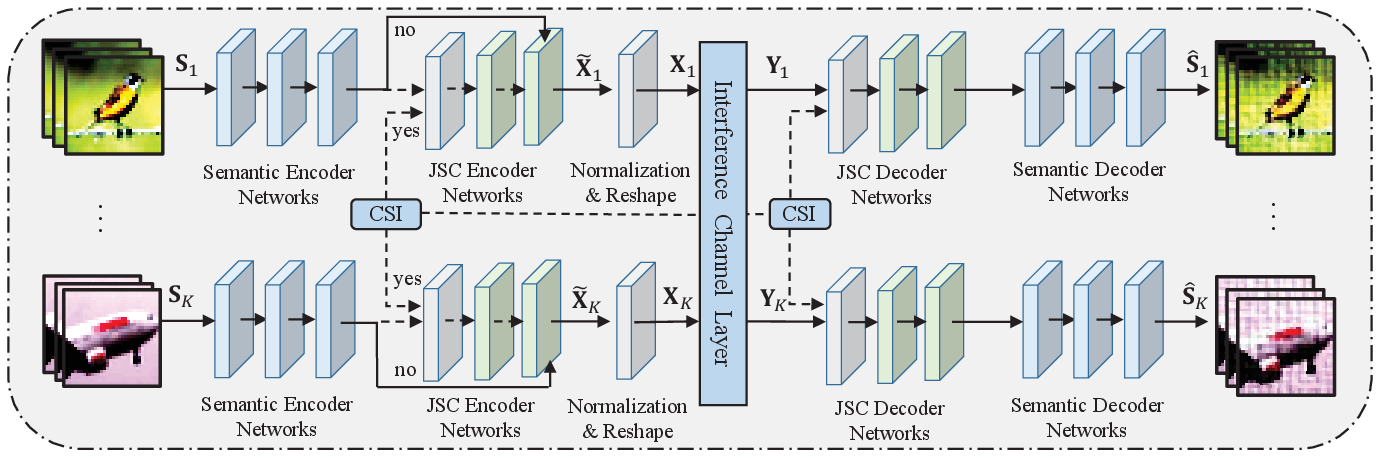}    
       \caption{The proposed neural network structure for IRSC.}
    \label{fig2}
\end{figure*}

\section{Interference Robust Semantic Communication}
To address problem \eqref{P1}, we design an interference-robust semantic communication system, named IRSC.
Specifically, we first propose to integrate CSI at the receiver only or at both the transmitter and receiver
ends, allowing IRSC to learn to adapt to channel interference.
Subsequently, we provide a detailed description of the model design.
Finally, we introduce the design of the loss function and the training strategy.

\subsection{CSI Integration for Enhancing Semantic Communications}
CSI plays a critical role in characterizing communication channels,
encompassing aspects such as fading and interference.
Having knowledge of CSI is imperative for optimizing the performance of communication systems.
In our work, we propose the integration of CSI into the IRSC system.
Specifically, we consider two different design options.
In the first case, where CSI is available at each receiver, we provide CSI, denoted as $\mathbf{H}$, along with the received signal $\mathbf{Y}_k$ to the JSC decoder.
$\mathbf{H}$ is initially transformed into a real-valued vector of length $2K^2 N_t N_r$ and then concatenated with $\mathbf{Y}_k$, which is similarly converted into a real-valued vector.
In the second scenario, where CSI is available at both the transmitter and receiver, the usage of $\mathbf{H}$ at the receiver follows the previously described approach.
At the transmitter, $\mathbf{H}$, which is converted into a real-valued vector with a length of $2K^2 N_t N_r$, is input to the JSC encoder together with the extracted semantic information to learn the features adapted to the interference channel.

\subsection{Model Design}
As shown in Fig. \ref{fig2},
the semantic encoder first learns semantic information from the inputs $\mathbf{S}_k$ and outputs the learned features.
It is noteworthy that the proposed IRSC is a universal framework, and the NN structure of the semantic encoder is not fixed.
For devices with limited computational and memory resources, we use a fully connected layer as the semantic encoder network to meet resource constraints.
Assuming abundant computational and memory resources, the semantic encoder can be composed of ResNet\cite{7780459}.
After passing through the semantic encoder network, we determine whether CSI is available at the transmitter.
If available, the CSI is fed to the JSC encoder along with the semantic information.
The first layer of the JSC encoder is a reshape layer that concatenates the semantic information with CSI and performs dimension transformation.
The second layer is a fully connected layer where the dimensions are reduced to the length of semantic information.
The third layer is also a fully connected layer, mapping the fused features to channel input symbols $\widetilde{\mathbf{X}}_k$.
If CSI is not available, the JSC encoder directly maps the semantic information to $\widetilde{\mathbf{X}}_k$, and in this case, the JSC encoder consists of only a single fully connected layer.
After that, $\widetilde{\mathbf{X}}_k$ is normalized and reshaped to $\mathbf{X}_k$.

In order to be able to jointly optimize the communication system in Fig. \ref{fig1} in an end-to-end manner, we model the interference channels as untrainable layers and incorporate them into the entire neural network architecture.
After the channel layer, $\mathbf{X}_k$ is converted to $\mathbf{Y}_k$.
$\mathbf{Y}_k$ and CSI are jointly input into the JSC decoder,
and its network structure is the same as the network structure of JSC encoder when CSI is available at the transmitter, which is also composed of three layers.
The semantic decoder network reverses the operations performed by the semantic encoder network and generates an estimate $\hat{\mathbf{S}}_k$ of the original image based on the output of the JSC decoder.
All the transmitters and receivers in IRSC are trained jointly, the loss is calculated at the end of the receiver and backpropagated to the transmitter, and the training parameters are updated simultaneously.

\subsection{Loss Function Design}

To jointly optimize the entire IRSC neural network,
we use the weighted combination of \eqref{P1} as the loss function, which can be written as
\begin{equation}
\begin{split}
\label{P2}
\mathcal{L}_{total} &= \underset{\bm{\alpha}_k, \bm{\beta}_k, \bm{\gamma}_k, \bm{\theta}_k} {\rm argmin}
\bigg\{ \sum^K_k  \omega_k \mathcal{L}_{MSE}(\bm{\alpha}_k, \bm{\beta}_k, \bm{\gamma}_k, \bm{\theta}_k) \bigg\}
\\&=\underset{\bm{\alpha}_k, \bm{\beta}_k, \bm{\gamma}_k, \bm{\theta}_k} {\rm argmin}  \bigg\{ \omega_1 \cdot \frac{1}{M} \sum^M_i (\mathbf{S}_{1,i}-\hat{\mathbf{S}}_{1,i})^2 + \cdots \\
& \quad +\omega_K \cdot \frac{1}{M} \sum^M_i (\mathbf{S}_{K,i}- \hat{\mathbf{S}}_{K,i})^2 \bigg\}, 
\end{split}
\end{equation}
where $\omega_k$ represents the weight assigned to the loss for user $k$.
In order to ensure equitable system performance among users, we implement a dynamic updating mechanism for $\omega_k$.
Specifically, the setting of $\omega_k$ is determined by the proportion of the corresponding user's contribution to the total loss in the previous iteration, which is given by
\begin{equation}
\label{xs}
\omega_{k}^{(t+1)} = \frac{\mathcal{L}_{MSE}(\bm{\alpha}_k^{(t)}, \bm{\beta}_k^{(t)}, \bm{\gamma}_k^{(t)}, \bm{\theta}_k^{(t)})}{\sum_k^K \mathcal{L}_{MSE}(\bm{\alpha}_k^{(t)}, \bm{\beta}_k^{(t)}, \bm{\gamma}_k^{(t)}, \bm{\theta}_k^{(t)})},
\end{equation}
where $t$ denotes the index of the current training epoch.
Initially, all weights are set to $1/K$.

\begin{algorithm}[t]
\caption{IRSC Training Algorithm.}
\label{alg:Framwork}
\small
\begin{algorithmic}[1] 
\REQUIRE
Initialize parameters, $\bm{\alpha}_k^{(0)}, \bm{\beta}_k^{(0)}, \bm{\gamma}_k^{(0)}, \bm{\theta}_k^{(0)}$,
\STATE \textbf{Input:} A batch of images $\mathbf{S}_k$, channel $\mathbf{H}$, epochs $T$.
\STATE \textbf{while} $t=1$ to $T$ \textbf{do}
\STATE \quad \textbf{if} CSI is only available at the receiver \textbf{then}
\STATE \quad \quad $\widetilde{\mathbf{X}}_k$ $\leftarrow$ $f_{\boldsymbol{\mathit{\beta}}_k}(f_{\boldsymbol{\mathit{\alpha}}_k}(\mathbf{S}_k))$.
\STATE \quad \textbf{else} CSI is available at both transmitter and receiver \textbf{then}
\STATE \quad \quad $\widetilde{\mathbf{X}}_k$ $\leftarrow$ $f_{\boldsymbol{\mathit{\beta}}_k}(f_{\boldsymbol{\mathit{\alpha}}_k}(\mathbf{S}_k), \mathbf{H})$.
\STATE \quad \textbf{end if}
\STATE \quad Power normalization.
\STATE \quad Transmit $\mathbf{X}_k$ over MIMO interference channel. 
\STATE \quad $\hat{\mathbf{S}}_k \leftarrow f_{\bm{\theta}_k}(f_{\bm{\gamma}_k}(\mathbf{Y}_k, \mathbf{H}))$.
\STATE \quad Compute loss $\mathcal{L}_{total}$ via \eqref{P2}.
\STATE \quad Update parameters simultaneously via SGD.
\STATE \quad $t\leftarrow t+1$.
\STATE \textbf{end while}
\STATE \textbf{Output:} $f_{\bm{\alpha}_k}(\cdot)$, $f_{\bm{\beta}_k}(\cdot)$, $f_{\bm{\gamma}_k}(\cdot)$, $f_{\bm{\theta}_k}(\cdot)$.
\end{algorithmic}
\end{algorithm}

\subsection{Training Strategy}

The goal of the aforementioned IRSC is to train an end-to-end model for transmitting images, particularly by leveraging CSI to enhance suitability for MIMO interference channels.
According to Fig. \ref{fig2}, the training algorithm for IRSC is outlined in Algorithm \ref{alg:Framwork}.
Initially, the NN parameters are initialized.
After extracting semantic information by the semantic encoder, the obtained semantic information is encoded, transmitted, and decoded alongside the CSI.
Subsequently, compute the loss and iteratively update the parameters using stochastic gradient descent (SGD).
Training persists until meeting termination criteria, such as reaching the maximum iteration count or observing no further reduction in the loss.

\emph{Complexity analysis:}
The computational complexity of the proposed IRSC depends on the neural network structure.
In the subsequent experiments, IRSC is implemented using fully connected layers.
The computational complexity of the fully connected layer is determined by the matrix multiplication operations.
Assuming the input feature dimension is $D_{\text{in}}$, the output feature dimension is $D_{\text{out}}$, and the batch size is $M$, the computational complexity of a single layer is $\mathcal{O}(M \times D_{\text{in}} \times D_{\text{out}})$.
Assuming there are $L$ layers of fully connected layer, the total computational complexity is $\mathcal{O}\left(\sum\limits_{l=1}\limits^L M \times D_{\text{in}}^{(l)}\times D_{\text{out}}^{(l)}\right)$.

\section{Experiments and Discussions}
In this section, several experiments are
provided to investigate the performance of the proposed IRSC scheme.

\subsection{Dataset and Parameter Setting}
We utilize the MNIST\cite{lecun1998gradient} dataset and the Fashion-MNIST\cite{FM} dataset for evaluation, as they are datasets commonly used in the field of machine learning and computer vision.
The MNIST dataset consists of handwritten digits from ``0'' to ``9'', while the Fashion-MNIST dataset contains $10$ different clothing categories.
They both contain $70,000$ grayscale images with a training set of $60,000$ examples and a test set of $10,000$ examples.
In the experiment, we set $K$ to $2$, indicating consideration for a two-user system.
Each user's transmitter and receiver are equipped with $2$ antennas respectively.
The neural network structures of the transmitter and receiver are listed in Table \ref{T1}.
Throughout the training process, we employ the Adam\cite{kingma2014adam} optimizer with 
learning rate $0.001$, betas of $0.9$ and $0.98$, batch size of $128$, and epoch of $200$.

\subsection{Evaluation Metric and Performance Baselines}

We use Structural Similarity (SSIM) as the performance metric\cite{1284395}, which measures the similarity of two images:
\begin{equation}
\label{ssim}
{\rm SSIM}(\mathbf{s},\hat{\mathbf{s}}) = \frac{(2\mu_s\mu_{\hat{s}}+C_1)(2\sigma_{s\hat{s}}+C_2)}{(\mu_s^2 +\mu_{\hat{s}}^2 +C_1)(\sigma_s^2 +\sigma_{\hat{s}}^2 +C_2)},
\end{equation}
where $\mu_s$ and $\sigma_s$ are the mean and the standard deviation for image ${\rm s}$;  $\sigma_{s\hat{s}}$ is the covariance of ${\rm s}$ and ${\rm \hat{s}}$;
$C_1$ and $C_2$ are constants to stabilize the division.
The SSIM index ranges from $-1$ to $1$.
A higher SSIM value suggests that the two images being compared are more similar.

\begin{table}[t]\scriptsize
\caption{Neural Network Structure of The
Transmitter and Receiver}
\centering
\label{T1}
\begin{tabular}{|c|c|c|}
\hline
                     & \textbf{Layer}                                                                                                                        & \textbf{Output} \\ \hline
\textbf{Transmitter} & \begin{tabular}[c]{@{}c@{}}Fully-connected Layer + ReLU\\ Reshape + Fully-connected Layer + ReLU\\ Fully-connected Layer\end{tabular} & 64              \\ \hline
\textbf{Receiver}    & \begin{tabular}[c]{@{}c@{}}Fully-connected Layer + ReLU\\ Fully-connected Layer +Tanh\end{tabular}                                    & 784             \\ \hline
\end{tabular}
\end{table}

To fully verify the effectiveness of the proposed IRSC, we consider the following baselines:

\textbf{Interference-Free scheme:}
The semantic information is transmitted to the receiver without noise and interference, which serves as the upper bound.

\textbf{Semi-Conventional scheme:}
Before semantic information is transmitted by the transmitter $k$, it undergoes preprocessing by the transmitting precoder.
At the receiving end, the signal received by the receiving antenna is first passed through a receiving filter, and then fed into the JSC decoder and semantic decoder for image recovery.
The signal of receiver $k$ after filtering can be written as
\begin{equation}
\label{semi}
\mathbf{Y}_k = \mathbf{U}_{k}^{\rm H} \mathbf{H}_{kk} \mathbf{V}_{k} \mathbf{X}_k
+ \sum_{j=1, j\neq k}^K \mathbf{U}_{k}^{\rm H} \mathbf{H}_{kj} \mathbf{V}_{j}\mathbf{X}_j + \mathbf{U}_{k}^{\rm H} \mathbf{N}_k,
\end{equation}
where $\mathbf{V}_{k} \in \mathbb{C}^{N_t\times N_r}$ and $\mathbf{U}_{k} \in \mathbb{C}^{N_r\times N_t}$ represent the precoding matrix and the receiving filtering matrix of user $k$, respectively.
For the design of $\mathbf{V}_k$ and $\mathbf{U}_k$, we adopt the method proposed in \cite{7222487}.

\textbf{CSI-Free scheme:}
The neural network structure of this scheme is the same as that of IRSC, and also adopts \eqref{P2} as the loss function.
The difference is that CSI is not integrated into the transmitter and receiver.
The purpose of this scheme is to verify the validity of the method proposed in Section III-A.


\subsection{Performance of IRSC}
\begin{figure}
       \centering   
       \includegraphics[width=0.85\linewidth]{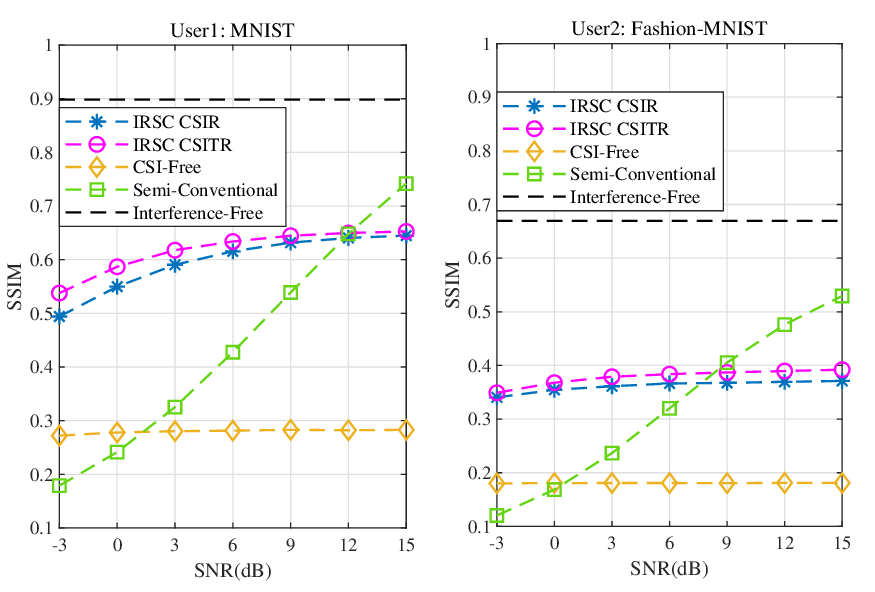}      
       \caption{Performance comparison of IRSC with other baseline schemes on the MNIST dataset and Fashion-MNIST dataset over MIMO interference channels.}
    \label{MF}
\end{figure}

\begin{figure}
       \centering   
       \includegraphics[width=0.9\linewidth]{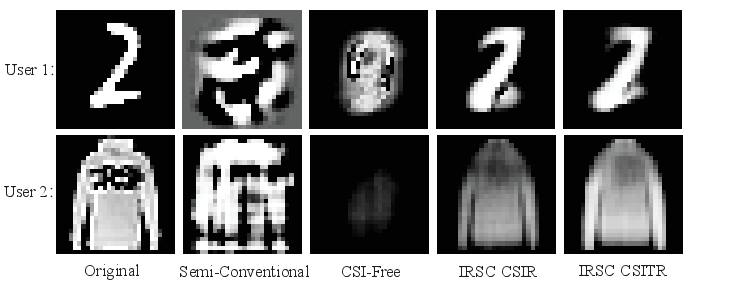}      
       \caption{Comparison of reconstructed images produced by various schemes in the low SNR regime: An Example at SNR = $0$dB.}
    \label{MAF}
\end{figure}

In the first experiment, the two users are assumed to have different message sets for transmission, where user $1$ aims to complete the MNNIST image transmission, while user $2$ aims to complete the Fashion-MNNIST image transmission.
Fig. \ref{MF} plots the performance of the proposed IRSC schemes and baselines over MIMO interference channels for different test SNR values.
We can observe that, whether it is user $1$ or user $2$, both the proposed IRSC CSIR (integrating CSI into receiver networks) scheme and IRSC CSITR (integrating CSI into transmitter and receiver networks) scheme outperform the CSI-Free scheme.
This indicates that the proposed schemes can effectively mitigate the impact of MIMO interference channels. 
In the high SNR regimes, the Semi-Conventional scheme approaches the upper bound more closely. 
However, this scheme requires the estimation of CSI at the receiver and then feeding it back to the transmitter.
In contrast, the proposed IRSC CSIR scheme does not require feedback CSI to the transmitter, and we also observe that it exhibits enhanced competitiveness in the low SNR regimes.
To illustrate more intuitively, the example reconstruction image with SNR = $0$ dB is shown in Fig. \ref{MAF}. Comparing IRSC CSIR with IRSC CSITR, we find they are achieving almost the same performance, implying that IRSC with CSI only at the receiver can handle the interference well.


\begin{figure}
       \centering   
       \includegraphics[width=0.85\linewidth]{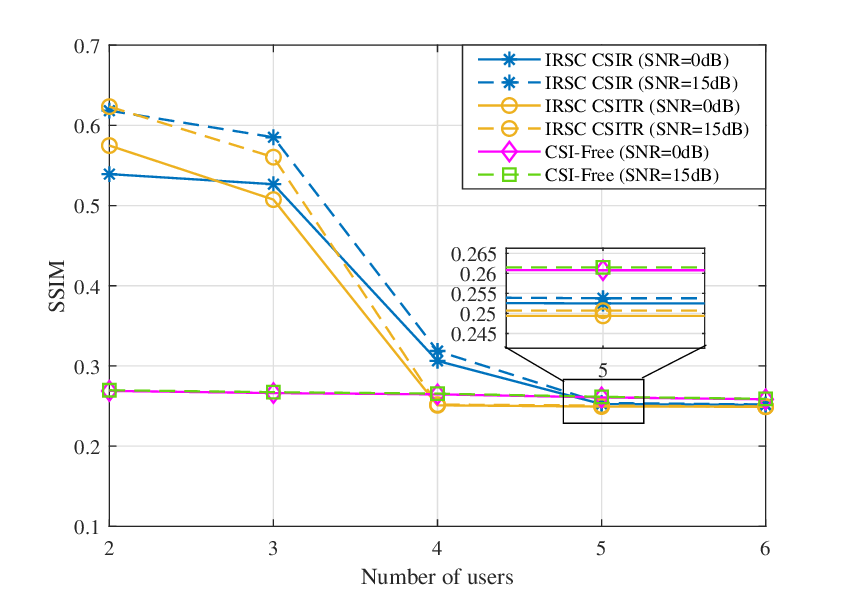}      
       \caption{SSIM comparisons versus the number of users over MIMO interference channels with test SNR of $0$dB and $15$dB.}
    \label{DYH}
\end{figure}

Next, we discuss the effect of different user numbers on the performance of the proposed IRSC schemes.
Assuming that all users are aiming to complete MNIST image transmission, we average the SSIM of all users as the performance of the algorithm.
The simulation results are shown in the Fig. \ref{DYH}.
It can be seen that in the case of fewer users, the proposed IRSC schemes can deal with interference more effectively, and they are easier to achieve good SSIM.
As the number of users increases, the interference problem becomes more complex, so SSIM of the proposed IRSC schemes decreases.
We also note that IRSC CSITR performs almost the same as IRSC CSIR.
Since IRSC CSIR does not require CSI feedback to the transmitter, it is preferred when handling interference.

\section{Conclusion}
In this paper, we evaluated the performance gains of semantic communication systems in interference scenarios.
Specifically, we proposed an interference-robust semantic communication scheme, named IRSC, over MIMO interference channels.
The IRSC scheme uses NNs for transceiver design and integrates CSI. 
In particular, we have considered two design options: IRSC CSIR, where CSI is input solely at the receiver end, and IRSC CSITR, where CSI is input at both transmitter and receiver ends.
We trained the transceivers by establishing a composite loss function and implemented a dynamic mechanism to enhance system fairness among users.
Experiments show that the proposed IRSC CSIR and IRSC CSITR can effectively mitigate interference and significantly outperform the baselines in the low SNR regime. Additionally, the IRSC CSIR does not require CSI feedback.



\bibliographystyle{IEEEtran} 
\bibliography{IEEEabrv,bib}

\end{document}